\documentclass[twocolumn,twoside,slac_two]{revtex4}
\usepackage{graphicx}
\usepackage{fancyhdr}
\usepackage{natbib}
\begin{document}

\title{Contribution to the Extragalactic Gamma-ray Background from the Cascades of Very-high Energy Gamma Rays}

%

\author{T. M. Venters}
\affiliation{NPP Fellow, GSFC, Greenbelt, MD 20771, USA}

\begin{abstract}
As very-high--energy photons propagate through the extragalactic background light (EBL), they interact with the soft photons and initiate electromagnetic cascades of lower energy photons and electrons. The collective intensity of a cosmological population emitting at very-high energies (VHE) will be attenuated at the highest energies through interactions with the EBL and enhanced at lower energies by the resulting cascade. We calculate the cascade radiation created by VHE photons produced by blazars and investigate the effects of cascades on the collective intensity of blazars and the resulting effects on the extragalactic gamma-ray background. We find that cascade radiation greatly enhances the collective intensity from blazars at high energies before turning over due to attenuation. The prominence of the resulting features depends on the blazar gamma-ray luminosity function, spectral index distribution, and the model of the EBL. We additionally calculate the cascade radiation from the distinct spectral sub-populations of blazars, BL Lacertae objects (BL Lacs) and flat-spectrum radio quasars (FSRQs), finding that the collective intensity of BL Lacs is considerably more enhanced by cascade radiation than that of the FSRQs due to their harder spectra. As such, studies of the blazar contribution to the EGRB by Fermi will have profound implications for the nature of the EBL, the evolution of blazars, and blazar spectra.
\end{abstract}

\maketitle

\thispagestyle{fancy}


\section{Introduction}

Since resolved blazars constitute the class of gamma-ray emitters with the largest number of identified members \citep{har99,lat09a}, unresolved blazars are expected to have a significant contribution to the extragalactic gamma-ray background (EGRB) at GeV energies \citep{sre98}. The amount of this contribution remains unclear, ranging from a few to $100\%$ of the EGRB \citep{ste93,sal94,ste96,kaz1997,chi98,muk99,muc00,der07,gio06,nar06,ino09} and depending not only the distribution of blazars with respect to gamma-ray luminosity and redshift (the blazar gamma-ray luminosity function, GLF), but also on the distribution of blazar spectral indices (SID) at GeV energies \citep{ste96,poh97,vp07,pv08}.  Additionally, photons with energies greater than ~20 GeV can interact with the lower energy photons of the extragalactic background light (IR/Opt./UV; EBL) as they propagate from high redshifts creating electromagnetic cascades of electrons, positrons, and photons. As blazars also constitute a cosmological population, the collective emission of unresolved blazars at gamma-ray energies will exhibit an absorption feature at the high end of the \emph{Fermi} LAT energy range.  The shape of the absorption feature is sensitive to the blazar GLF and the EBL \citep{vpr09}. The created cascade emission from absorption can enhance the collective blazar intensity at lower energies with the amount of enhancement being sensitive to blazar spectra at very-high energies \citep{ven09}.

\section{Contribution to the EGRB from Cascades of VHE Photons from Blazars}

The contribution to the EGRB due to blazars can be viewed as the superposition of the collective intensity of \emph{intrinsic} blazar spectra and the contribution from the cascade radiation from the interactions of very-high--energy photons from blazars with the EBL:
\begin{equation}
I_E(E_0) = I^i_E(E_0) + I^c_E(E_0)\,,
\end{equation}
where the intensity, $I_E(E_0)$, is given in units of photons per unit energy per unit time per unit area per unit solid angle emitted at observer-frame energy, $E_0$. The collective intensity of \emph{intrinsic} blazar spectra including attenuation by the EBL is given by \citep[for derivation, see][]{vpr09}:
\begin{widetext}
\begin{equation}
I^i_E(E_0)=\!\! \frac{c}{H_0} \frac{1}{4\pi E_f^2} \! \int_{-\infty}^{\infty} \!\!\!\! d\alpha\,p_L(\alpha)\!\!\left(\frac{E_0}{E_f}\right)^{-\alpha}\!\!\!\! \int_{z=0}^{z_{\rm max}} \!\!\!\!\!\! dz'\,\frac{e^{-\tau(E_0,z')}}{(1+z')^{\alpha}[\Omega_{\Lambda}+\Omega_m(1+z')^3]^{1/2}} \!\! \int_{L_{\gamma,{\rm min}}}^{L_{\gamma,{\rm max}}} \!\!\!\!\!\! dL_{\gamma}L_{\gamma}\rho_{\gamma},
\end{equation}
\end{widetext}
where $E_f$ is some fiducial \emph{observer} frame energy (taken to be $100$ MeV), \linebreak $L_{\gamma} = E_{f}'^2dN(E_{f}')/dtdE$ is the isotropic gamma-ray luminosity of a blazar at the fiducial \emph{rest} frame energy, $E_f' = E_f(1+z')$, $\rho_{\gamma} = d^2N/dL_{\gamma}dV_{\rm com}$ is the blazar GLF, $\tau(E_0,z)$ is the optical depth as a function of \emph{observed} photon energy and source redshift, $p_L(\alpha) = dN/d\alpha$ is the SID of gamma-ray blazars corrected for measurement error and sample bias \citep{vpr09}, $L_{\gamma,{\rm max}} = 4\pi d_L^2 (\alpha-1)(1+z)^{\alpha-2} E_f F_{\gamma,{\rm min}}$, $F_{\gamma,{\rm min}}$ is the sensitivity of the survey above the fiducial observer frame energy\footnote{For EGRET, the sensitivity above $100$ MeV is $10^{-7}$ $\mbox{photons } {\rm cm}^{-2} {\rm s}^{-1}$. For  \emph{Fermi} LAT, the one-year sensitivity above $100$ MeV is $2 \times 10^{-9}$ $\mbox{photons } {\rm cm}^{-2} {\rm s}^{-1}$.}.  In deriving the above equation, we have assumed the standard $\Lambda$CDM cosmology and that blazar energy spectra can be described as single power laws defined by the spectral index, $\alpha$ ($dN_{\gamma}/dE_{\gamma} \propto E_{\gamma}^{-\alpha}$).

Making the dependencies explicit, the cascade intensity is given by
\begin{equation}\label{eqn-cascadeintens}
I^c_E = \frac{d^4N^c}{dtdAd\Omega dE}\,,
\end{equation}
where $dN^c/dE$ is the spectrum of cascade photons due to pair production and Inverse Compton scattering \citep[for derivation, see][]{ven09}:
\begin{widetext}
\begin{equation}\label{eqn-cascadespec}
\frac{dN^c}{dE_0}(E_0) = \! \int_0^{z_{\rm max}} \!\!\! \int_{E_{p,{\rm min}}}^{E_{p,{\rm max}}} (1+z) \frac{d^2N_\gamma}{dzdE_p} P(f;E_p,z) \left[\frac{dN_{\Gamma_1}(E_0(1+z))}{dE}+\frac{dN_{\Gamma_2}(E_0(1+z))}{dE}\right] e^{-\tau(E_0,z)} dE_p dz,
\end{equation}
\end{widetext}
where $dN_\Gamma/dE$ is the spectrum of Inverse Compton scattered radiation per electron of Lorentz factor, $\Gamma$, $P(f;E_p,z)$ is the probability that the pair production interaction at a given redshift, $z$, of a primary photon of energy $E_p$ will produce electron-type particles of energies $E_{e1} = f\times E_p$ and $E_{e2} = (1-f)\times E_p$, $\Gamma_1 = f\times E_p/mc^2$, $\Gamma_2 = (1-f)\times E_p/mc^2$, and $d^2N_\gamma/dzdE_p$ is the continuous spectrum of photons undergoing pair production interactions as a function of redshift and primary energy.  Numerical integration of Eq. \ref{eqn-cascadespec} is both complicated and computationally expensive, and thus, we make use of a Monte Carlo propagation code, \emph{Cascata} \citep{ven09}, which propagates photons from known spectra and redshift distributions and calculates the cascade spectrum of each photon.

In determining the cascade contribution to the EGRB from VHE photons from blazars, primary photon energies are randomly generated according to the \emph{absorbed} EGRB intensity (in observer's quantities), which is the differential amount of radiation that has been ``absorbed'' by the soft photon background: 
\begin{widetext}
\begin{equation}\label{eqn-abs}
I^a_E(E_0)
=\!\! \frac{c}{H_0} \frac{1}{4\pi E_f^2} \int_{-\infty}^{\infty} \!\!\! d\alpha\,p_L(\alpha) \!\! \left(\frac{E_0}{E_f}\right)^{-\alpha}\!\!\! \int_{z=0}^{z_{\rm max}} \! dz\,\frac{(1-e^{-\tau(E_0,z)})}{(1+z)^{\alpha}[\Omega_{\Lambda}+\Omega_m(1+z)^3]^{1/2}} \int_{L_{\gamma,{\rm min}}}^{L_{\gamma,{\rm max}}} \!\!\!\!\! dL_{\gamma}L_{\gamma}\rho_{\gamma}\,.
\end{equation}
\end{widetext}
This intensity is interpreted as the radiation that creates electromagnetic cascades.  As such, it will also be used to normalize the cascade spectra in the final step.

The \emph{absorbed} EGRB intensity is also used to generate the redshifts of the sources of the VHE photons by differentiating with respect to redshift: 
\begin{widetext}
\begin{equation}\label{eqn-absdz}
\frac{dI^a_E}{dz}(E_0) = \frac{c}{H_0}\frac{1-e^{-\tau(E_0,z)}}{4\pi E_f^2[\Omega_{\Lambda}+\Omega_m(1+z)^3]^{1/2}} \int_{-\infty}^{\infty} \! d\alpha\,p_L(\alpha)(1+z)^{-\alpha}\left(\frac{E_0}{E_f}\right)^{-\alpha} \!\!\! \int_{L_{\gamma,{\rm min}}}^{L_{\gamma,{\rm max}}} \!\!\!\!\! dL_{\gamma}L_{\gamma}\rho_{\gamma}\,.
\end{equation}
\end{widetext}
The created photons are then propagated until the redshift of interaction, randomly generated from the probability of interaction, $p = 1-\exp[-\tau(E_0,z)]$.

\section{Results}

\begin{figure*}[t]
\centering
\includegraphics[width=120mm]{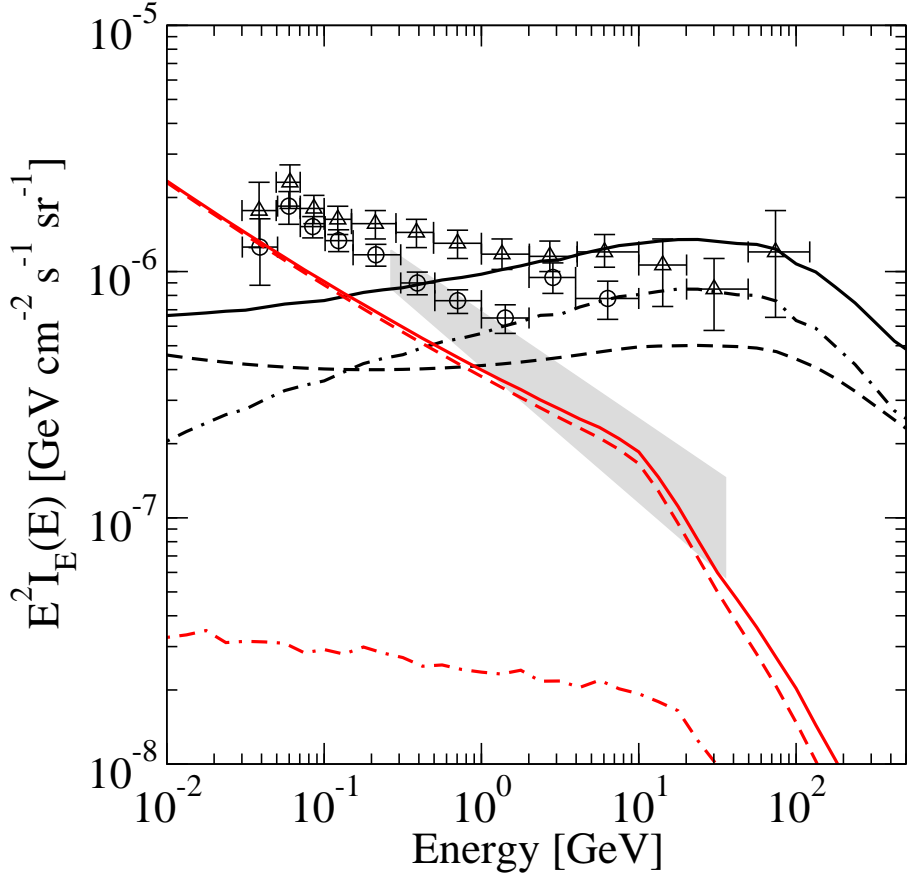}
\caption{Collective intensity of blazars including cascade radiation computed assuming separate SIDs and GLFs for BL Lacs (black; $\alpha_0 = 1.97$, $\sigma_0 = 0.175$; luminosity-dependent density evolution) and FSRQs (red; $\alpha_0 = 2.37$, $\sigma_0 = 0.14$; pure luminosity evolution). \emph{Solid}: Total collective intensity including both the collective intensity from intrinsic blazar spectra (including absorption) and cascade radiation. \emph{Dashed}: Collective intensity assuming only absorption. \emph{Dot-dashed}: Cascade intensity assuming the Stecker EBL model. \emph{Gray band}: Preliminary \emph{Fermi} EGRB (presented at the 31$^{\rm st}$ International Cosmic Ray Conference by M. Ackermann). \emph{Data points:} EGRET EGRB (\emph{triangles}: \citet{sre98}; \emph{circles}: \citet{str04})}\label{FSRQ_BLL_Stecker_Fermi_color-f1}
\end{figure*}

\begin{figure*}[t]
\centering
\includegraphics[width=120mm]{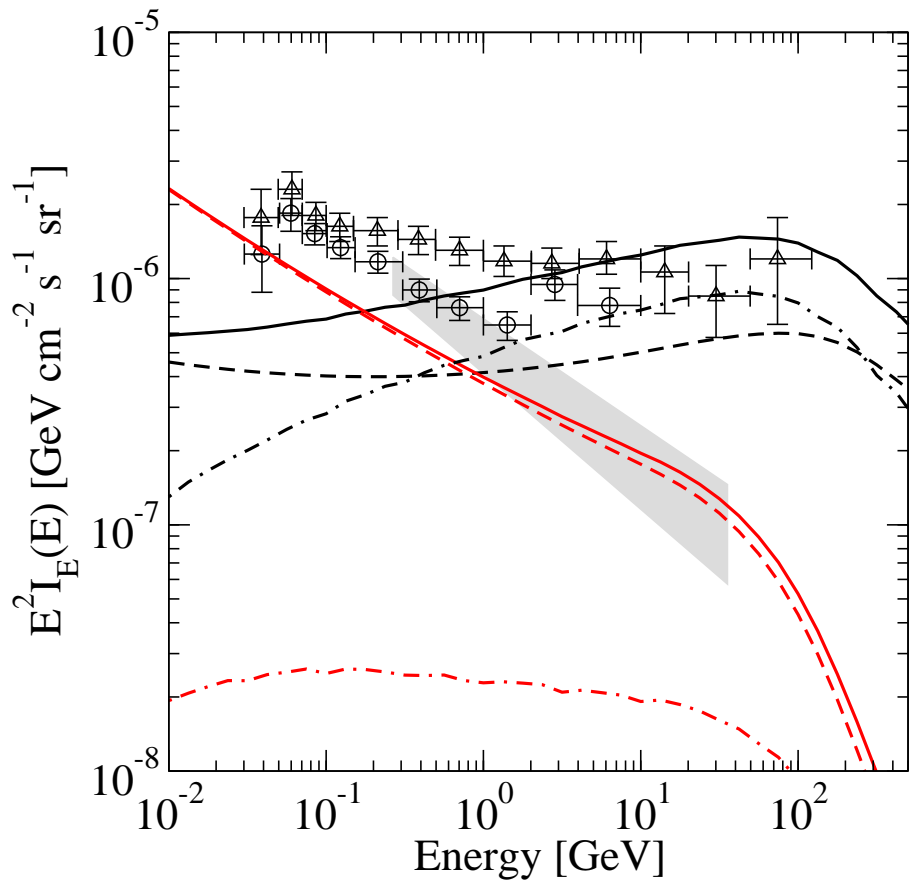}
\caption{Same as in Figure \ref{FSRQ_BLL_Stecker_Fermi_color-f1} except for the Gilmore EBL model.}\label{FSRQ_BLL_Gilmore_Fermi_color-f2}
\end{figure*}

We determine the blazar contributions to the EGRB including cascade radiation assuming that the two blazar sub-populations, BL Lacertae-like objects (BL Lacs) and flat spectrum radio quasars (FSRQs), form separate populations with respect to emission and evolution. We propagate photons from distinct GLFs and SIDs for the two blazar sub-populations. We assume BL Lacs to follow a luminosity-dependent density evolution model (LDDE) and FSRQs to follow the pure luminosity evolution model (PLE) \citep[for more information about both models, see][]{nar06}. We determine the parameters of the SIDs for each blazar sub-population by applying the maximum-likelihood analysis of \citet{vp07} to the spectral indices of the blazars listed in the \emph{Fermi} LAT Bright AGN Sample \citep{lat09}.  Results for the \citet{ste06} EBL model and the \citet{gil09} EBL model are plotted in Figures \ref{FSRQ_BLL_Stecker_Fermi_color-f1} and \ref{FSRQ_BLL_Gilmore_Fermi_color-f2}, respectively.

In both EBL models, the effect of the cascade radiation is to flatten the overall collective intensity from blazars.  Below $10$ GeV, the cascade radiation for the two models is roughly comparable, though there is slightly more cascade radiation at the lowest energies for the Stecker model.  Above $10$ GeV, there is more cascade radiation for the Gilmore model than there is for the Stecker model due to the enhanced UV absorption in the Stecker model.  The resulting high-energy absorption feature in the overall spectrum is much more prominent in the Stecker case than in the Gilmore case owing to the higher UV background in the Stecker EBL model\footnote{Since the pair production cross section as a function of the center-of-momentum energy peaks at twice the electron mass squared, one would expect that gamma-ray photons of energies $\sim$ tens of GeV are most likely to interact with UV background photons. Thus, unsurprisingly, models with high UV backgrounds will result in more suppression at high energies.}. This is most especially true for the FSRQs, which follow the PLE model of evolution, resulting in more high-energy photons being produced at higher redshifts than in the LDDE model \citep{vpr09}.

The hard spectra of BL Lacs cause the resulting cascade radiation to be considerable relative to that of the collective intensity from their intrinsic spectra even though their collective radiation tends to be concentrated at low redshifts in the LDDE model\footnote{However, note that we did not take into account possible breaks in the spectra.}. In fact, the total intensity from BL Lacs could dominate the EGRB at high energies and even overproduce it, though it doesn't explain the lower energy intensity. By contrast, the relatively steep spectra of FSRQs are not conducive to much cascade radiation even though their collective radiation is concentrated more at high redshifts; hence, most of the collective intensity attributed to FSRQs occurs at lower energies.  Thus, while the BL Lacs (including cascades) dominate the collective blazar intensity at the highest energies, the FSRQs dominate at the lowest energies.

It should be noted that we have neglected the possible effects of propagation through intervening magnetic fields, which we will examine in a future publication.

\acknowledgements{We acknowledge enlightening feedback from Angela Olinto, Vasiliki Pavlidou, Kostas Tassis, Floyd Stecker, Joel Primack, and Marjorie Corcoran.
This work was supported in part by the Kavli Institute for Cosmological Physics at the University of Chicago through grants NSF PHY-0114422 and NSF PHY-0551142 and an endowment from the Kavli Foundation and its founder Fred Kavli. T. M. V. was also supported by the NSF Graduate Research Fellowship Program and by an appointment to the NASA Postdoctoral Program at the Goddard Space Flight Center, administered by Oak Ridge Associated Universities through a contract with NASA.}


\end{document}